\documentclass[12pt]{article}
\usepackage{graphicx,amssymb,amsfonts,amsmath,epsfig,color}

\makeatletter
\DeclareSymbolFont{AMSb}{U}{msb}{m}{n}
\DeclareSymbolFontAlphabet{\mathbb}{AMSb}
\renewcommand{\section}{\@startsection{section}{1}{\z@}%
                                    {-7ex \@plus -1ex \@minus -.2ex}%
                                    {2.5ex \@plus.2ex}%
                                    {\normalfont\large\scshape\centering}}
\renewcommand{\subsection}{\@startsection{subsection}{2}{\z@}%
                                       {-5ex \@plus -1ex \@minus -.2ex}%
                                       {1.5ex \@plus.2ex}%
                                       {\normalfont\normalsize\scshape}}
\renewcommand{\subsubsection}{\@startsection{subsubsection}{3}{\z@}%
                                       {-5ex \@plus -1ex \@minus -.2ex}%
                                       {1.5ex \@plus.2ex}%
                                       {\normalfont\normalsize\scshape}}

\renewcommand\@seccntformat[1]{\ignorespaces\csname #1name\endcsname\space
                               \csname the#1\endcsname.\quad}   
\newdimen\captionmargin
\newdimen\captionindent
\newdimen\captionwidth
\newcommand{\captionfont}{\slshape}
\newcommand\@captionlabel[1]{\textsc{#1:}\space}
\long\def\@makecaption#1#2{%
  \vskip\abovecaptionskip
  \captionwidth\hsize
  \advance\captionwidth -2\captionmargin
  \sbox\@tempboxa{\@captionlabel{#1}\captionfont #2}%
  \ifdim \wd\@tempboxa >\captionwidth
    \ifdim\captionindent>\z@
      \advance\captionwidth -\captionindent
      \hskip\captionindent
    \fi
    \hskip\captionmargin
    \parbox[t]{\captionwidth}{\leavevmode\hskip-\captionindent
      \@captionlabel{#1}\captionfont #2}%
  \else
    \global \@minipagefalse
    \hb@xt@\hsize{\hfil\box\@tempboxa\hfil}%
  \fi
  \vskip\belowcaptionskip}
\def\eqnarray{%
   \stepcounter{equation}%
   \def\@currentlabel{\p@equation\theequation}%
   \global\@eqnswtrue
   \m@th
   \global\@eqcnt\z@
   \tabskip\@centering
   \let\\\@eqncr
   $$\everycr{}\halign to\displaywidth\bgroup
       \hskip\@centering$\displaystyle\tabskip\z@skip{##}$\@eqnsel
      &\global\@eqcnt\@ne$\;\hfil{##}$\hfil
      &\global\@eqcnt\tw@$\;\displaystyle{##}$\hfil\tabskip\@centering
      &\global\@eqcnt\thr@@ \hb@xt@\z@\bgroup\hss##\egroup
         \tabskip\z@skip
      \cr}
\ifcase \@ptsize
\or
\or
\fi
  \divide\@tempdima\baselineskip
  \@tempcnta=\@tempdima
\makeatother
\begin{document}

\renewcommand{\theequation}{\arabic{section}.\arabic{equation}}
\renewcommand{\thefigure}{\arabic{figure}}
\newcommand{\gapprox}{%
\mathrel{%
\setbox0=\hbox{$>$}\raise0.6ex\copy0\kern-\wd0\lower0.65ex\hbox{$\sim$}}}
\textwidth 165mm \textheight 220mm \topmargin 0pt \oddsidemargin 2mm
\def\ib{{\bar \imath}}
\def\jb{{\bar \jmath}}

\newcommand{\ft}[2]{{\textstyle\frac{#1}{#2}}}
\newcommand{\be}{\begin{equation}}
\newcommand{\ee}{\end{equation}}
\newcommand{\bea}{\begin{eqnarray}}
\newcommand{\eea}{\end{eqnarray}}
\newcommand{\Identity}{{1\!\rm l}}
\newcommand{\cx}{\overset{\circ}{x}_2}
\def\CN{$\mathcal{N}$}
\def\CH{$\mathcal{H}$}
\def\hg{\hat{g}}
\newcommand{\bref}[1]{(\ref{#1})}
\def\espai{\;\;\;\;\;\;}
\def\zespai{\;\;\;\;}
\def\avall{\vspace{0.5cm}}
\newtheorem{theorem}{Theorem}
\newtheorem{acknowledgement}{Acknowledgment}
\newtheorem{algorithm}{Algorithm}
\newtheorem{axiom}{Axiom}
\newtheorem{case}{Case}
\newtheorem{claim}{Claim}
\newtheorem{conclusion}{Conclusion}
\newtheorem{condition}{Condition}
\newtheorem{conjecture}{Conjecture}
\newtheorem{corollary}{Corollary}
\newtheorem{criterion}{Criterion}
\newtheorem{defi}{Definition}
\newtheorem{example}{Example}
\newtheorem{exercise}{Exercise}
\newtheorem{lemma}{Lemma}
\newtheorem{notation}{Notation}
\newtheorem{problem}{Problem}
\newtheorem{prop}{Proposition}
\newtheorem{rem}{{\it Remark}}
\newtheorem{solution}{Solution}
\newtheorem{summary}{Summary}
\numberwithin{equation}{section}
\newenvironment{pf}[1][Proof]{\noindent{\it {#1.}} }{\ \rule{0.5em}{0.5em}}
\newenvironment{ex}[1][Example]{\noindent{\it {#1.}}}

\thispagestyle{empty}


\begin{center}

{\LARGE\scshape On the interior of (Quantum) Black Holes
\par}
\vskip15mm

\textsc{R. Torres}\footnote{E-mail: ramon.torres-herrera@upc.edu}
\par\bigskip
{\em
Department of Applied Physics, UPC, Barcelona, Spain.}\\[.1cm]

\end{center}


\begin{abstract}
Different approaches to quantum gravity conclude that
black holes may possess an inner horizon, in addition to the (quantum corrected) outer `Schwarzschild' horizon. In this paper we assume the existence of this inner horizon and explain the physical process that might lead to the tunneling of particles through it. It is shown that the tunneling would produce a flux of particles with a spectrum that deviates from the pure thermal one. Under the appropriate approximation the extremely high temperature of this horizon is calculated for an improved quantum black hole. It is argued that the flux of particles tunneled through the horizons affects the dynamics of the black hole interior leading to an endogenous instability.
\end{abstract}






\setcounter{equation}{0}

\section{Introduction}

The discovery that black holes emit radiation had a big impact on the scientific community. The celebrated pioneering work on this subject was performed by Hawking in 1975 \cite{Haw75} who showed, based on results of quantum field theory on a fixed curved background (Schwarzschild's solution), that black holes emit a thermal spectrum of particles from their event horizon.
The heuristic picture most commonly proposed as an explanation of this effect is that of pair creation near the horizon of the black hole and the corresponding tunneling of particles in which one of the components of the pair is swallowed by the black hole and the other escapes.
This picture led Parikh and Wilczek \cite{P&W} to propose a method for studying Hawking radiation from the Schwarzschild horizon by explicitly considering the tunneling of particles through it. Furthermore, their method took into account the back-reaction effect of the radiation on the black hole thanks to the requirement of energy conservation and showed that new terms appear in the distribution function which deviate it from pure thermal emission, i.e., the standard Boltzmann distribution.

Of course, this picture is incomplete since,
in order to describe the last stages of black hole evaporation, one should take into account quantum gravity effects. The possibility of studying the radiation from the outer horizon of quantum corrected black holes is now feasible from different approaches to Quantum Gravity \cite{B&RIS}\cite{Modes}\cite{Amel}\cite{Nicol}.
Sometimes a strict thermal evolution has been imposed on the quantum black hole by estimating Hawking's energy flux directly from Stefan-Boltzmann's law. However, it is also possible to study more accurately the radiation from quantum black holes  by following the approach of Parikh and Wilczek. For example, in \cite{RPO} the tunneling of particles through the outer horizon has been studied by using an effective quantum spacetime \cite{B&RIS}
based on the Quantum Einstein Gravity approach.

On the other hand, the possibility that black holes could have an inner horizon seems nowadays plausible since the results from different frameworks \cite{B&RIS}\cite{Modes}\cite{Amel}\cite{Nicol} point in this direction. However, while there exists a vast amount of work devoted to the properties of the outer horizon, the properties of this inner horizon remain, in comparison, relatively unknown.
It seems, therefore, natural at this moment to speculate about the properties of this horizon and its consequences on the inner dynamics of the black hole.
This is the aim of this Letter in which the possibility of tunneling from the inner horizon is studied (specifically for the solution found in \cite{B&RIS}) and the physical process behind it is explained.
Moreover, guided by the well-known existence of classical solutions possessing an inner horizon instability under the perturbation of external fields (from which, the Reissner-Nordstr\"{o}m solution is the paramount example), the stability of the inner horizon of a quantum corrected solution is checked. In particular, we are interested not only in the influence of external fields, but in whether the flux of energy tunneled through the black hole horizons could have consequences on its own stability.

The Letter has been divided as follows. Section \ref{secISS} introduces the solution for the quantum black hole (the \textit{improved Schwarzschild spacetime}) and its main properties. In section \ref{secstab} the stability of the solution is checked under the action of a test distribution of noninteracting massless particles. Section \ref{sectun} analyzes the tunneling of particles through the inner horizon of the improved black hole.
This allows us, in section \ref{secdist}, to evaluate the spectral distribution and temperature of the emitted particles. The flow of energy through the inner horizon is found in section \ref{secFlow} and a model for the evolution of an evaporating quantum black hole is then treated in section \ref{secBR}. The stability of the evaporating model under the flow of energy from its horizons is analyzed in section \ref{secerevstab}. Finally, the results are discussed in section \ref{seccon}.

\section{Improved Schwarzschild solution}\label{secISS}
In \cite{B&RIS} Bonanno and Reuter found an effective spacetime for a quantum black hole by using the idea of the Wilsonian renormalization group \cite{Wilson} in order to study quantum effects in the Schwarzschild spacetime. Specifically, they obtained a \textit{renormalization group improvement} of the Schwarzschild metric based upon a scale dependent Newton constant $G$ obtained from the exact renormalization group equation for gravity \cite{Reuter} describing the scale dependence of the effective average action \cite{Wett}\cite{R&W}.
The solution can be written as
\begin{equation}\label{RGISch}
ds^2=- f(R) dt_S^2+f(R)^{-1} dR^2+ R^2 d\Omega^2.
\end{equation}
where
\begin{equation}
f(R)=1-\frac{2 G(R) M}{R}
\end{equation}
with
\begin{equation}
G(R)=\frac{G_0 R^3}{R^3+\tilde{\omega} G_0 (R+\gamma G_0 M)} \label{GR}
\end{equation}
and where $G_0$ is Newton's universal gravitational constant, $M$ is the mass measured by an observer at infinity and $\tilde{\omega}$ and $\gamma$ are constants coming from the non-perturbative renormalization group theory and from an appropriate cutoff identification, respectively.
In \cite{B&RIS}\cite{B&RIV} it is argued that the preferred value for $\gamma$ is $\gamma=9/2$.
On the other hand, $\tilde \omega$ can be found by comparison with the the standard perturbative quantization of Einstein's gravity (see \cite{Dono} and references therein). It can be deduced that its precise value is $\tilde \omega=167/(30\pi)$, but the properties of the solution do not rely on its precise value as long as it is strictly positive.

The horizons in this solution can be found by solving $f(R)=0$.
The number of positive real solutions to this equation correspond to the positive real solutions of a cubic equation and depends on the sign of its discriminant or, equivalently, on whether the mass is bigger, equal or smaller than a critical value $M_{cr}$. In general, the critical value takes the form
\begin{equation}
M_{cr}=a(\gamma) \sqrt{\frac{\tilde{\omega}}{G_0}}=a(\gamma) \sqrt{\tilde{\omega}} m_p \sim \sqrt{\tilde{\omega}} m_p,
\end{equation}
where $m_p$ is Planck's mass and the function $a(\gamma)$ has, in general, an involved expression that, for reasonable values of $\gamma$ satisfies $a(\gamma)\sim 1$. In particular, the preferred value $\gamma=9/2$ provide us with
\[
M_{cr}=\frac{1}{24} \sqrt{\frac{1}{2} (2819+85 \sqrt{1105})}\sqrt{\frac{\tilde\omega}{G_0}}\simeq 2.21 \sqrt{\tilde{\omega}} m_p \simeq 2.94 m_p.
\]

\begin{figure}
\includegraphics[scale=.8]{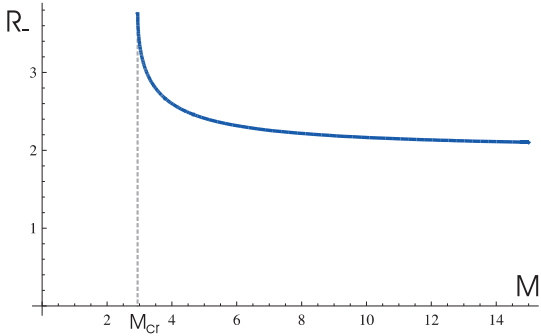}
\caption{\label{figR-} $R_-(M)$ is plotted in Planck units for masses around the critical mass. A calculation shows that $R_-(M=M_{cr})\simeq 3.772$ while $R_-(M\rightarrow\infty)\simeq 1.997$.}
\end{figure}

If $M<M_{cr}$ the equation has not positive real solutions, so that there are not horizons. If $M=M_{cr}$ there is only one positive real solution to the cubic equation. Finally, if $M>M_{cr}$ then the equation has two positive real solutions $\{R_-,R_+\}$ satisfying $R_-<R_+$.
The outer solution $R_+$ can be considered as the \textit{improved Schwarzschild horizon}, i.e., the Schwarzschild horizon when quantum modifications are taken into account. On the other hand, the inner solution $R_-$ represents a novelty with regard to the classical solution. It is a monotonically decreasing function of $M$ defined for masses non-smaller than the critical mass (see figure \ref{figR-}) that from its maximum value $R_+(M_{cr})\ (\simeq 3.772 \sqrt{G_0})$ tends asymptotically towards the value $R_{-min}=\sqrt{G_0 \gamma \tilde{\omega}/2}$.

\begin{figure}
\includegraphics[scale=.8]{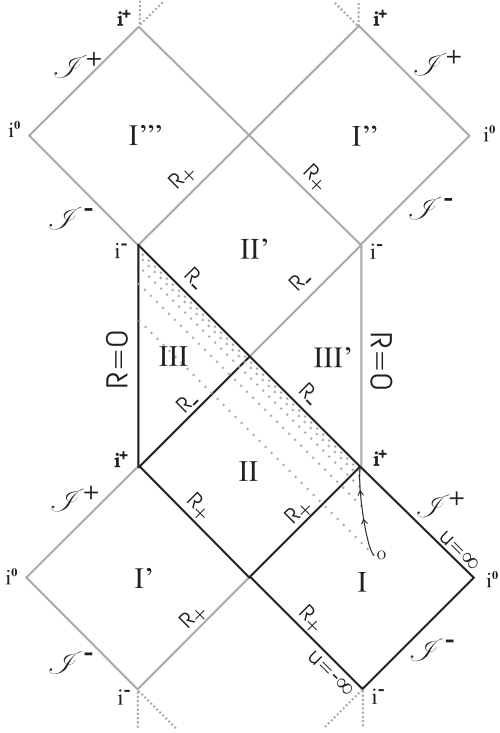}
\caption{\label{figstat} A Penrose diagram corresponding to the case $M>M_{cr}$. The region drawn using a solid black line (I-II-III) correspond to the zone defined by the solution in Eddington-Finkelstein-like coordinates (\ref{ScEF}) with the null coordinate going from $u=-\infty$ to $u=\infty$. The regions drawn in grey correspond to extensions of this solution. Penrose's classical instability argument \cite{Penrose} is schematically shown: An eternal observer `O' emits radiation (dashed lines) at equal intervals from the asymptotically flat region I towards $R=0$. As it approaches its timelike infinite $i^+$ the radiation piles up at the inner horizon $R_-$ ($u=\infty$), which is an instable surface of infinite blueshift.}
\end{figure}

The maximally extended spacetime for this solution in the case $M>M_{cr}$ resembles the Reissner-Nordstr\"{o}m maximally extended spacetime in the case $M>|Q|$. A Penrose diagram of the improved black hole for this case has been depicted in figure \ref{figstat}.
Note that the usual $R=0$ singularity in the classical Schwarzschild solution does not exist in the improved solution \cite{B&RIS}\cite{impcoll}. It is also important to remark for later purposes that, from a classical point of view and as can be directly checked from fig.\ref{figstat}, a photon in region II that follows the ingoing direction towards region III must reach $R=0$.

In order to interpret the physical meaning of this solution let us suppose that it has been generated by an effective matter fluid in such a way that the coupled gravity-matter system satisfies Einstein's equations $G_{\mu\nu}=8\pi G_0 T_{\mu\nu}$.
Consider now a radially moving observer with an arbitrary 4-velocity $\mathbf{v}$ and an orthonormal basis  $\{ \mathbf{v},\mathbf{n}, \mbox{\boldmath{$\omega$}}_\theta,
\mbox{\boldmath{$\omega$}}_\varphi \}$ such that
\mbox{\boldmath{$\omega$}}$_\theta \equiv R\ d\theta$,
\mbox{\boldmath{$\omega$}}$_{\varphi}\equiv R\ sin\theta \
d{\varphi}$ and $ {\bf n} $ is a space-like 1-form.
Any radially moving observer can write the vacuum energy-momentum tensor as an anisotropic fluid
\begin{eqnarray}\label{TV}
\mathbf{T}_V =& \varrho_V \mathbf{v} \otimes
\mathbf{v} + p_V \mathbf{n}\otimes \mathbf{n}+ p_{\bot}
(\mbox{\boldmath{$\omega$}}_\theta \otimes \mbox{\boldmath{$\omega$}}_\theta +
\mbox{\boldmath{$\omega$}}_\varphi \otimes \mbox{\boldmath{$\omega$}}_\varphi)\ , \label{VIV}
\end{eqnarray}
where $\varrho_V$ is the vacuum energy density, $p_V$ is the vacuum normal pressure and $p_{\bot }$ is the vacuum tangential pressure.
By using the field equations, one can obtain their explicit expressions:
\begin{eqnarray}
\varrho_V &=& \frac{M G'}{4 \pi G_0 R^2} =-p_V, \label{varrho}\\
p_{\bot } &=& -\frac{M G''}{8 \pi G_0 R},\nonumber
\end{eqnarray}
where $G'$ and $G''$ are, respectively, the first and second derivatives of $G$ with respect to $R$.

\section{Testing the stability of the solution}\label{secstab}

It has long been known that, for the Reissner-Nordstr\"{o}m solution with $M>|Q|$, slight deviations from its initial conditions produce large effects at their inner horizon (see, for example, \cite{S&P}\cite{C&H}). More precisely, physical quantities associated with the perturbations, such as the energy density measured by a free-falling observer, diverge at their inner horizon, so that the horizon is unstable to time-dependent perturbations. Since the same classical arguments that led to the discovery of the Reissner-Nordstr\"{o}m instability apply for our solution (see the caption to figure \ref{figstat}) it would be interesting to test whether the inner horizon of the quantum corrected solution
may also be unstable.

A well-known preliminary test of stability consists in considering a simple model involving a test distribution of noninteracting massless particles falling inside the black hole (see, for example, \cite{Poisson}).
In order to perform this test, let us first write the improved Schwarzschild's metric (\ref{RGISch}) in terms of ingoing Eddington-Finkelstein-like coordinates $\{u,R,\theta,\varphi\}$, where
\[
u=t_S+\int^R \frac{dR'}{1-2 G(R') M/R'}\ ,
\]
as
\begin{equation}\label{ScEF}
ds^2=-\left(1-\frac{2 G(R) M}{R}\right) du^2+2 du dR + R^2 d\Omega^2.
\end{equation}
Let us recall that the existence of an inner horizon requires $M\geq M_{cr} $ and remark that we are interested in the stability of the inner horizon that, in these coordinates, correspond to the lightlike surface $\{u=+\infty, R=R_-\}$ (see figure \ref{figstat}).

\begin{figure}
\includegraphics[scale=.9]{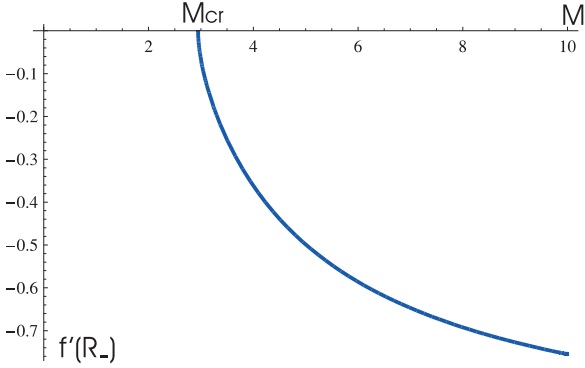}
\caption{\label{figgmenos} A plot of the function $f'(R_-)$. It is negative for $M>M_{cr}$ and null for $M=M_{cr}$.}
\end{figure}

The test incoming massless particles can be described by considering an energy-momentum tensor for the radiation with the form
\begin{equation}
\mathbf{\mathcal{R}} = \mu\,
\mathbf{l} \otimes \mathbf{l}, \label{RIV}
\end{equation}
where  $\vec l=-\partial/\partial R$ is a radial light-like 4-vector pointing in the
direction of the future directed ingoing radiation.
In this way, a radially moving observer measures a density of radiation:
\begin{equation}\label{densrad}
\rho_{Rad}=\mathcal{R}_{\alpha\beta} v^\alpha v^\beta=\mu (v^u)^2,
\end{equation}
where $v^u$ is the $u$-component of the observer normalized four-vector $\mathbf{v}$. In order to explicitly compute this measured density of radiation, let us consider the particular case of a free-falling observer. In this case the four vector satisfies the geodesic equation
\[
\frac{d v^\nu}{d \tau}+\Gamma^\nu_{\alpha\beta} v^\alpha v^\beta=0,
\]
where $\tau$ is the observer proper time. This provide us with
\[
\frac{d v^u}{d \tau}+\frac{f'}{2} (v^u)^2=0,
\]
where $f'$ is the derivative of $f$ with respect to $R$.
Integrating it we get
\begin{equation}\label{vu}
v^u= c \exp\left(\frac{-1}{2} \int f' du \right),
\end{equation}
where $c$ is an integration constant. Close to the inner horizon we have
\begin{equation}\label{kneq0}
f'\simeq f'(R_-)=2 \kappa_-,
\end{equation}
where $\kappa_-$ is the surface gravity of the inner horizon.
For our solution, $\kappa_-<0$ if $M>M_{cr}$ and $\kappa_-=0$ if $M=M_{cr}$ (see figure \ref{figgmenos}.)

Let us first consider the case $M>M_{cr}$.
Using (\ref{kneq0}) in (\ref{vu}), the density of radiation (\ref{densrad}) measured by the free-falling observer near the inner horizon becomes
\begin{equation}
\rho_{Rad}\simeq c^2 \mu \exp(- 2 \kappa_- u  ).
\end{equation}
In this way, if, as expected (see section \ref{secFlow}), $\mu$ decayed following a power law $u^{-p}$ ($p>0$) as $u$ approaches infinity, we would have close to the inner horizon
\begin{equation}
\rho_{Rad}\sim u^{-p} \exp(- 2 \kappa_- u  ) \ \ \mbox{as}\ \ u\rightarrow\infty.
\end{equation}
Therefore, the measured energy density would diverge as the observer reaches the inner horizon, what indicates the presence of an instability in the inner horizon for the case $M>M_{cr}$.

In case $M=M_{cr}$ and, in this way, $\kappa_-= 0$ the behaviour of (\ref{vu}) as $u$ tends to infinity and $f'$ approaches zero along the timelike geodesic has to be checked. In this case we should replace (\ref{kneq0}) with $f'(R)\simeq f''(R_-) (R-R_-)$ (where, for our improved solution, $f''(R_-)\simeq 0.1294/G_0>0$).
So that (\ref{densrad}) can be written approximately as
\begin{equation}\label{densradk0}
\rho_{Rad} \simeq c^2 \mu \exp\left(- f''(R_-) \int (R(u)-R_-) du\right).
\end{equation}
We would now need the approximate behaviour of the geodesic $R(u)$ close to the inner horizon. In order to get it, we take into account that
$v^\alpha v_\alpha=-1$ can be written, using (\ref{vu}), as
\begin{equation}
\frac{dR}{du}=
\frac{1}{2} \left(f - \frac{1}{c^2} \exp\left(\int f' du \right) \right).
\end{equation}

The solution of this integro-differential equation with the boundary condition $R(u\rightarrow\infty)=R_-$
can be approximated for $u\sim \infty$ by
\begin{equation}
R(u)= R_- - \frac{4}{f''(R_-) u}+ O(u^{-2}).
\end{equation}
Equipped with this result we can now compute the measured density of radiation (\ref{densradk0}) close to the inner horizon as
\[
\rho_{Rad}\sim \mu u^4.
\]
Therefore, if $\mu$ tended to zero following a power law $u^{-p}$ ($p>0$) as $u$ approaches infinity we would have
\begin{equation}
\rho_{Rad} \sim u^{4-p}.
\end{equation}
In this way, the instability of the inner horizon under a test radiation for the case $M=M_{cr} \Leftrightarrow \kappa_-=0$ seems only to appear if $0<p<4$ or, in other words, if the decay of the test radiation as the inner horizon is approached is not too fast. Thus, the analysis of the possible sources of radiation becomes crucial for the conclusions about the stability in this case.

\section{Tunneling through the inner horizon}\label{sectun}

In order to analyze the possibility of tunneling through the inner horizon let us consider that a pair of photons is created in region II (fig.\ref{figstat}), where the 2-spheres are closed trapped surfaces.
A pair of test photons would be classically forced to move inwards until reaching $R=0$. However, for non-test photons energy conservation modifies this picture near the inner horizon since the positive energy photon produced in the pair could `tunnel' the inner horizon and move \emph{outwards} in region III (or III').
This possibility, that would seem impossible in view of figure \ref{figstat}, is feasible because
energy conservation implies that, as the black hole mass would be reduced in such a process, the inner horizon would expand (backreaction) provoking the tunneling.
(A sketch of the process, taking into account the backreaction, is
shown in figure \ref{figbackr}).

Let us now put these ideas into practice by considering an improved black hole satisfying $M>M_{cr}$. First, we will rewrite the improved Schwarzschild's solution in Painlev\'e-like coordinates \cite{Pain} so as to have coordinates which are not singular at the horizons. In order to do this it suffices to introduce a new coordinate $t$ replacing the Schwarzschild-like time $t_S$ such that $t=t_S+h(R)$ and fix h(R) by demanding the constant time slices to be flat. In this way one gets:
\begin{equation}
ds^2=-\left(1-\frac{2 G(R) M}{R}\right) dt^2+2 \sqrt{\frac{2 G(R) M}{R}} dt dR+ dR^2 + R^2 d\Omega^2,
\end{equation}
where $R$ can now take the values $0<R<\infty$.
In these coordinates the radial null geodesics describing the evolution of \emph{test} massless particles are given by
\begin{equation}\label{geodtest}
\frac{dR}{dt}=\pm 1-\sqrt{\frac{2 G(R) M}{R}}
\end{equation}
with the upper (lower) sign corresponding to outgoing (ingoing, respectively) geodesics.

In \cite{K&W}\cite{P&W}\cite{I&Y} it was found that, when a self-gravitating shell of energy $E$ travels in
a spacetime characterized by an ADM mass $M$,
the geometry outside the shell is also characterized by $M$, but energy conservation implies that
the geometry inside the shell is characterized by $M-E$. It was also found that the shell
then moves on the geodesics given by the interior line element.
In this way, according to (\ref{geodtest}), one expects a shell of energy $E$ to satisfy the evolution equation
\begin{equation}\label{geodshell}
\frac{dR}{dt}=\pm 1-\sqrt{\frac{2 G(R;E) (M-E)}{R}}
\end{equation}
where $G(R;E)$ is the function $G(R)$ with $M$ replaced by $M-E$\footnote{Note that, from now on, the nomenclature `$;E$' means `$M$ should be replaced by $M-E$' is used for all the functions appearing in this Letter whose dependence on $E$ is explicited.}.

Let us consider pair production occurring near the inner horizon with the positive energy particle tunneling outwards.
The semiclassical emission rate will be given by $\Gamma \sim \exp\{-2 \mbox{Im} S\}$, where $S$ is the particle action. Therefore, we have to compute the imaginary part of the action for an
outgoing positive energy particle which crosses the horizon $R_-$ inwards from $R_{in}$ in the trapped region II to $R_{out}$ in III beneath the inner horizon.
\begin{equation}
\mbox{Im} S=\mbox{Im} \int_{R_{in}}^{R_{out}} p_R dR= \mbox{Im}  \int_{R_{in}}^{R_{out}}  \int_{0}^{p_R} dp'_R dR.
\end{equation}
Using Hamilton's equation $\dot{R}=+dH/dp_R\rfloor_R$ and $H=M-E'$, this can be written with the help of (\ref{geodshell}) as
\begin{eqnarray}\label{imS}
\mbox{Im} S&=& \mbox{Im} \int_{M}^{M-E} \int_{R_{in}}^{R_{out}} \frac{dR}{\dot R}  dH=\nonumber\\
&=&\mbox{Im} \int_{0}^{E} \int_{R_{in}}^{R_{out}} \frac{dR}{1-\sqrt{\frac{2 G(R;E') (M-E')}{R}}} (-dE').
\end{eqnarray}
If we define the functions $f(R;E')$ and $g(R;E')$ such that
\begin{displaymath}
f(R;E')\equiv 1-\frac{2 G(R;E') (M-E')}{R}=(R-R_-(E')) g(R;E'),
\end{displaymath}
where $R_-(E')$ is the position of the inner horizon when $M$ is replaced by $M-E'$ and $g$ satisfies
\begin{equation}
g(R_-;E')=
\frac{\partial f(R;E')}{\partial R}\rfloor_{R=R_-(E')}.
\end{equation}
Note, from our comments in section \ref{secstab} (now with $M\rightarrow M-E'$), that $g(R_-;E')$ is negative for $M-E'>M_{cr}$, null for $M-E'=M_{cr}$ and it does not exist for $M-E'<M_{cr}$.

Then, the integral for the variable $R$ in (\ref{imS}) can be performed by
following Feynman's prescription $E'\rightarrow E'-i \epsilon$ (so that the pole is in the upper-half \textit{R}-plane) and taking into account that energy conservation implies that a particle in the trapped region II, just close to the inner horizon, tunnels towards an expanding horizon ($R_{in}<R_{out}$). In this way, one gets
\[
 \int_{R_{in}}^{R_{out}} \frac{dR}{1-\sqrt{\frac{2 G(R;E') (M-E')}{R}}} = \frac{2 \pi i }{g(R_-;E')}.
\]

So that we can write (\ref{imS}) as
\begin{equation}\label{partchan}
\mbox{Im} S=- \int_{0}^{E}  \frac{2 \pi}{g(R_-;E')} dE'\ ,
\end{equation}

obtaining for the semiclassical rate
\begin{equation}\label{emprob}
\Gamma\sim e^{-2 \mbox{\scriptsize Im} S }=\exp\left(4\pi \int_{0}^{E}  \frac{dE'}{g(R_-;E')} \right).
\end{equation}

\section{Temperature and spectral distribution}\label{secdist}

\begin{figure}
\includegraphics[scale=.9]{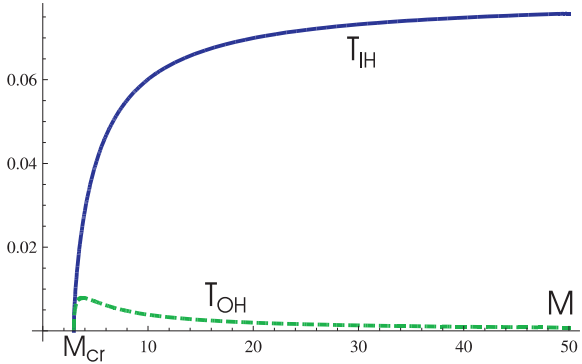}
\caption{\label{figtemp} A plot of the temperatures at the inner (solid line) and outer (dashed line) horizons as a function of the black hole mass.}
\end{figure}

If quadratic terms were neglected
we could develop Im $S$ up to first order in $E$ as
\[
\mbox{Im} S \simeq \frac{2 \pi}{g(R_-,0)} E
\]
obtaining a thermal radiation at the inner horizon (IH) of the quantum black hole ($\Gamma \sim \exp\{-E/T_{IH}\}$) with temperature
\begin{equation}\label{TQBH}
T_{IH}=-\frac{g(R_-,0)}{4 \pi}=-\frac{1}{4 \pi}\left. \frac{\partial f}{\partial R}\right\rfloor_{R=R_-}.
\end{equation}
This coincides with the expected temperature obtained by computing it as $T=|\kappa|/(2 \pi)$ (coming, for instance, from an Euclidean continuation of the near horizon geometry).
By checking the properties of $f'(R_-)$ (fig.\ref{figgmenos}) one sees that the temperature assigned to the interior horizon will be a monotonically increasing function of $M$ that is null for black holes with masses equal to the critical mass  and that approaches, as the mass increases, the value
\begin{equation}\label{Tmax}
\lim_{M \rightarrow \infty} T_{int}=\frac{1}{\sqrt{2 \pi^2 \gamma \tilde{\omega} G}}=\frac{T_p}{\sqrt{2 \pi^2 \gamma \tilde{\omega}}},
\end{equation}
where $T_p$ is Planck's temperature. It is interesting to note the satisfactory result that, according to (\ref{Tmax}), the temperature of the black hole's inner horizon never exceeds Planck`s temperature, but is bounded below the value $T_{int}\simeq 7.97\cdot 10^{-2}\ T_p$. Nevertheless, the inner temperature will be extremely high for macroscopic black holes. In figure \ref{figtemp} this temperature is compared with the temperature of the outer horizon (that can be found in \cite{B&RIS}\cite{RPO}).

Notwithstanding these results about the temperature of the black hole, it is important to remark that the higher order terms in $E$, neglected in order to obtain (\ref{TQBH}), imply a deviation from pure thermal emission.
Let us recall that the standard thermal distribution for photons is
\begin{displaymath}
<n(E)>_{Therm.}=\frac{1}{\exp(\frac{E}{T})-1},
\end{displaymath}
with $0\leq E<\infty$, what taking into account the temperature of the interior horizon (\ref{TQBH}) can also be written as
\begin{equation}\label{nStand}
<n(E)>_{Therm.}=\frac{1}{\exp(\frac{- 4 \pi E}{g(R_-;0)})-1}.
\end{equation}

\begin{figure}
\includegraphics[scale=.9]{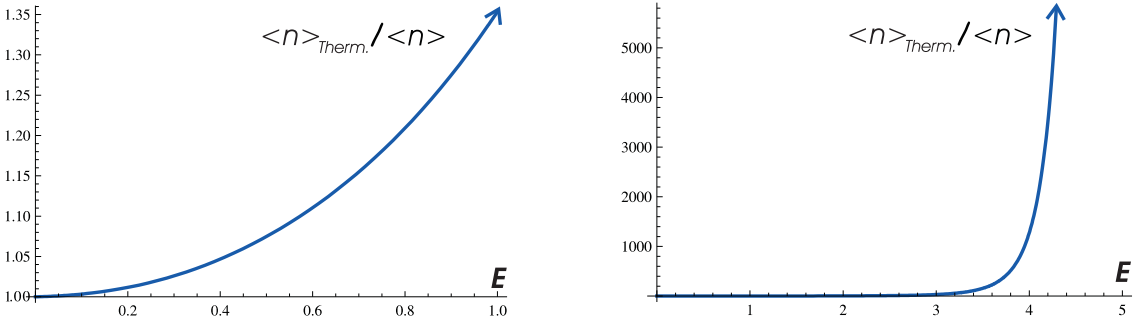}
\caption{\label{figdist} Graphics of the ratio $<n(E)>_{Therm.}/<n(E)>$ (for the arbitrary value $M=10$) showing that the distribution of the emitted particles deviates from a thermal distribution. The left-hand graphic shows that the deviation is smaller for extremely low energy particles, while the right-hand graphic hints that the deviation is unbounded as the energy of the emitted particle approaches its maximum value $E=M-M_{cr}$.}
\end{figure}

On the other hand, if we consider the full consequences of energy conservation, the distribution function for the emission of photons
can be written as (see \cite{K&K} --correcting the result in \cite{K&W}--)
\begin{displaymath}
<n(E)>=\frac{1}{\exp \left(2 \mbox{Im} S \right)-1}.
\end{displaymath}
What for our quantum corrected solution becomes in the inner horizon
\begin{equation}\label{nE}
<n(E)>=\frac{1}{\exp \left(- 4\pi \int_{0}^{E}  \frac{dE'}{g(R_-;E')} \right)
-1},
\end{equation}
with the additional requirement that, according to the properties of $g(R_-;E')$,
the energy of the emitted particles must satisfy $E\leq M-M_{cr}$.
This is very interesting since it imposes energy conservation by forbidding the emitted quantum to carry more energy than the black hole mass. In fact, this can be taken as an indication that a thermal spectrum, which would contain a tail of arbitrarily high energies, can not provide us with the correct spectrum.
If one compares the standard thermal distribution (\ref{nStand}) with the distribution (\ref{nE}) one sees that they are only similar for particles with extremely low energy. However, the thermal distribution always provides a bigger mean number of radiated particles and the ratio $<n(E)>_{Therm.}/<n(E)>$ increases without bound as the energy of the emitted particle approaches its maximum value $E=M-M_{cr}$ (see figure \ref{figdist}).

\section{Flow of energy}\label{secFlow}

\begin{figure}
\includegraphics[scale=.9]{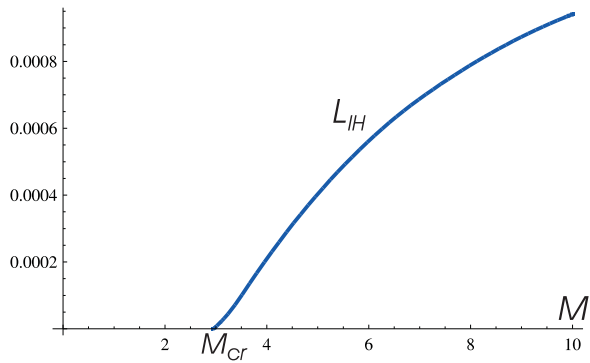}
\caption{\label{lumi10} The flow of energy tunneled through the inner horizon as a function of the black hole mass. The figure only shows the behaviour for small masses. For macroscopic masses $L_{IH}\simeq 1.66 \cdot 10^{-3}$.}
\end{figure}

Using (\ref{nE}), the flow of positive energy directed towards the asymptotically flat region I'' (fig.\ref{figstat})  due to the tunneled particles at the inner horizon can be written approximately as \cite{B&D}\cite{FN-S}
\begin{eqnarray}
L_{IH}(M)\simeq \frac{1}{2 \pi}\int_0^{M-M_{cr}} <n(E)>  E dE \nonumber\\
=\frac{1}{2 \pi}\int_0^{M-M_{cr}} \frac{E}{\exp\left(-4\pi \int_{0}^{E}  \frac{dE'}{g(R_-;E')} \right)-1} dE,\label{lumi}
\end{eqnarray}
where we are considering that the probability that a photon moving outwards in region III could be backscattered towards $R=0$ is small enough and, as shown in the previous section, we are compelled to take into account in the integration limits that
the maximum energy of a radiated particle could be
$M-M_{cr}$.
It can be checked that
the flow of energy is null if $M=M_{cr}$ and it is an increasing function of the black hole mass (see figure \ref{lumi10}) that very soon approaches its asymptotic value $L_{IH}(M\ggg M_{cr})\simeq 1.66 \cdot 10^{-3}$.
On the other hand, since for masses around the critical mass one gets $g(R_-;E)\simeq a (M-M_{cr}-E)^{1/2}$, where $a\simeq 0.2514$ in Planck's units, (\ref{lumi}) tells us that the behaviour of the flow of energy for black holes with  masses of this small magnitude will satisfy
\begin{equation}\label{Linap}
L_{IH}(M)= \frac{5 a}{48 \pi^2} (M-M_{cr})^{3/2}+ \mathcal{O}(M-M_{cr})^2.
\end{equation}

\begin{figure}
\includegraphics[scale=.9]{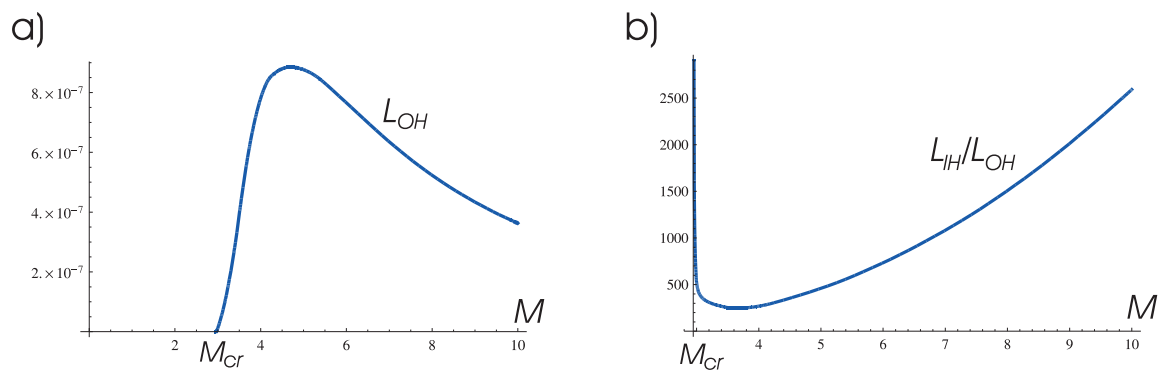}
\caption{\label{figcomp} To the left, the flow of energy tunneled through the outer horizon as a function of the black hole mass for small masses. (For bigger masses it becomes smaller as the considered mass increases. Eventually, the flow of energy tends asymptotically to zero). To the right, a comparison between the flow of energy from the inner and outer horizons.}
\end{figure}

We could now compare this flow of energy with the expected flow due to the tunneling through the outer horizon. The calculation taking into account energy conservation and the backscattered radiation can be found in \cite{RPO}. The flow has the form shown in figure \ref{figcomp}a and it is much smaller than the flow in the inner horizon
(figure \ref{figcomp}b).
On the other hand, let us comment for later purposes that the flow of energy in the outer horizon (OH) for black hole masses around the critical mass satisfies \cite{RPO}
\begin{equation}\label{Loutap}
L_{OH}\simeq k (M-Mcr)^{7/2},
\end{equation}
where $k$ is a constant.

\section{Modeling the backreaction}\label{secBR}

The improved solution (\ref{ScEF}) does not reflect the back-reaction associated to the lost of mass due to the tunneling effect. However, we can modelize the mass lost taking into account that, whenever a pair of virtual particles is created, when the particle with positive energy moves in the outgoing direction its companion, with negative energy, falls into the black hole thus reducing its mass. Let us recall that this flow of negative energy particles in the ingoing direction is due to the tunneling in the inner horizon and, in a lesser extent, to the tunneling in the outer horizon.
In this way, if we consider negative energy massless particles following ingoing null geodesics $u=$constant, the mass of the black hole becomes a decreasing function $M(u)$. The metric which incorporates the effect of the decreasing BH mass due to the ingoing null radiation
is (\ref{ScEF}) with $M$ replaced by $M(u)$, i.e., it corresponds to an \textit{improved} ingoing Vaidya solution \cite{B&RIV}
\begin{equation}\label{Vaid}
ds^2=-\left(1-\frac{2 G(R; M(u)) M(u)}{R}\right) du^2+2 du dR + R^2 d\Omega^2.
\end{equation}

\begin{figure}
\includegraphics[scale=1]{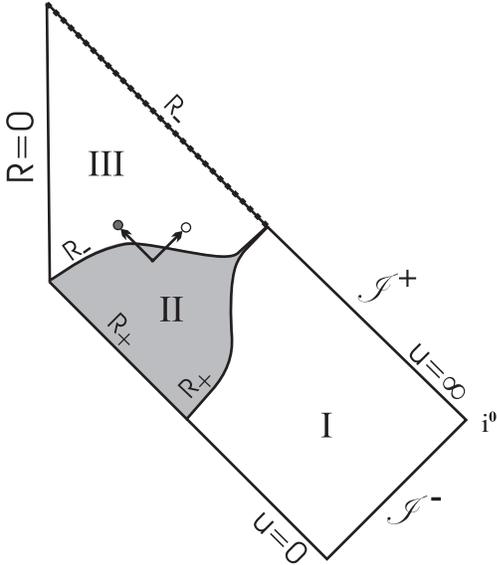}
\caption{\label{figbackr} A Penrose diagram of an evaporating black hole as described by the solution (\ref{Vaid}) for $0\leq u <\infty$ when $M(u=0)=M_0$. The backreaction to the tunneling of particles in the inner and outer horizon is reflected in that the inner horizon expands while the outer horizon shrinks. The tunneling of particles through the inner horizon has been schematically shown: A pair is created in region II. The negative energy particle (darker circle) falls towards $R=0$ while the positive energy particle (lighter circle) \textit{tunnels} outwards and then follows the outgoing direction in region III. On the other hand, the dashed line on the inner horizon ($u=\infty$) represents our ignorance on the resolution of the instability.}
\end{figure}

On the other hand, the flux of negative energy particles directed towards the black hole equals the flux of outgoing radiated particles and, therefore,
\begin{equation}\label{difM}
\frac{d M(u)}{du}=- L_{Total}(M(u)),
\end{equation}
where $L_{Total}=L_{IH}+L_{OH}$ takes into account the flow of negative energy from the inner and the outer horizons. From (\ref{difM}) and the positivity of $L_{Total}$ for $M>M_{cr}$, it seems clear that the mass of the black hole has to decrease until it reaches a mass close to the critical mass. When $M\sim M_{cr}$,
taking into account that (comparing (\ref{Linap}) with (\ref{Loutap})) $L_{Total}\simeq L_{IH}$, one can solve (\ref{difM}) to get the evolution of the mass
\begin{equation}\label{massevol}
M(u)\simeq M_{cr}+\frac{M_0-M_{cr}}{[1+\tilde{a} (M_0-M_{cr})^{1/2} u ]^2},
\end{equation}
where $\tilde{a}\equiv 5 a/(96 \pi^2)$ and we are using the initial condition $M(0)=M_0\ (\gtrsim M_{cr})$.
We see that the mass quickly approaches the value $M_{cr}$ for big enough values of $u$, i.e., as the inner horizon is approached. A Penrose diagram of the portion described by the solution (\ref{Vaid}) with the mass (\ref{massevol}) in the range $0\leq u<\infty$ is shown in figure \ref{figbackr}.
On the other hand, using (\ref{difM}) for (\ref{massevol}), we see that the flow of energy near the inner horizon ($u=\infty$) decays as
\begin{equation}\label{Lu}
L_{Total}\sim u^{-3}.
\end{equation}

\section{Revisiting stability}\label{secerevstab}

Any radially moving observer can write the effective energy-momentum tensor for (\ref{Vaid}) as
\begin{equation}
\mathbf{T} =\mathbf{T}_V+
\mathbf{\mathcal{R}}\ ,\label{TIVR}
\end{equation}
where $\mathbf{T}_V$ is the part of the
energy-momentum tensor corresponding to the vacuum energy density and pressures that we found previously (\ref{TV})-(\ref{varrho}) and $\mathbf{\mathcal{R}}$ is the
radiative part satisfying
$\mathbf{\mathcal{R}} = \mu\,
\mathbf{l} \otimes \mathbf{l}$,
where, as previously, $\vec l=-\partial/\partial R$ is a radial light-like 4-vector pointing in the
direction of the future directed radiation.
By using the field equations, one can obtain the explicit expression for the density of negative energy radiation
\begin{equation}
\mu =-\left(1-\frac{\gamma \tilde{\omega} G_0^2 M}{R^3+\tilde{\omega} G_0 (R+\gamma G_0 M)}\right) \left(\frac{G}{G_0}\right) \frac{ L_{Total}}{4 \pi R^2},\label{phi2}
\end{equation}
that corresponds to the quantum corrected version of the expression for the density of radiation $|\mu_{Stand.}|=L_{Total}/(4 \pi R^2)$.
In this way, (\ref{Lu}) inform us that the density of radiation satisfies near the inner horizon ($u=\infty$) satisfies
\begin{equation}\label{muIH}
\mu \sim L_{Total} \sim u^{-3}.
\end{equation}
In section \ref{secstab} we showed that stability of the inner horizon ($u=\infty$) requires $M=M_{cr}$ and a decay of $\mu$ as $u^{-p}$ with $p\geq 4$.
When black hole evaporation is taken into account, we have seen that, indeed, the mass reaches the critical mass on the inner horizon ($u=\infty$), but taking into account the decay of $\mu$ (\ref{muIH}), one has to conclude that an instability in the inner horizon ($u=\infty$) due to the tunneling in the inner horizon seems unavoidable.

As a final remark, note that if we \textit{forget} about the tunneling in the inner horizon and its corresponding flow of energy and consider just the tunneling in the outer horizon, then, solving (\ref{difM}) using (\ref{Loutap}), the mass for $u\sim\infty$ would have satisfied $M(u)-M_{cr}\sim u^{-2/5}$. Consequently, the density of radiation (\ref{phi2}) would have been $\mu\sim L\sim u^{-7/5}$ and again we see that $p=7/5<4$. In this way, the conclusion of the existence of an endogenous instability in the inner horizon (now due to the tunneling from the outer horizon) is again obtained.

\section{Conclusions}\label{seccon}

The only route to test existing theories about the interior of black holes is to examine their own physical and mathematical consistency under extreme conditions. In this Letter we have partially tried this by using a semiclassical tunneling approach to the improved solution appearing in \cite{B&RIS}. However, it can be argued that the results would be applicable to other quantum corrected effective solutions possessing an inner horizon and coming from different approaches to quantum gravity \cite{Modes}\cite{Amel}\cite{Nicol}. In particular, one would expect that the tunneling process, the explained physical mechanism behind it and the general properties for the inner horizon derived from it should be independent of the used framework. On the other hand, the test of stability has been performed for arbitrary $f(R)$, so that its results are expected to be also generalizable, although one must be aware that this is just a preliminary test.

Specifically, we have seen that the existence of an inner horizon could lead to the tunnel of particles through it and that the mechanism behind this tunneling is energy conservation: Once a pair is created in region II (fig.\ref{figstat}) close to the inner horizon, the negative energy particle can fall into the black hole what causes a decrease in the black hole mass and the subsequent expansion of the inner horizon, which is the necessary ingredient in order to the positive energy particle to scape outwards.

A temperature can be assigned to the distribution of tunneled particles under a rough approximation. This temperature turns out to be much bigger than any other known temperature in the current epoch of the Universe. Nevertheless, we have also seen that the distribution of tunneled particles deviates significantly from a thermal one as particles with high energy are considered.
On the other hand, the predicted flow of energy through the inner horizon has been shown to be bounded, but much bigger than the corresponding flow for particles tunneled through the outer horizon.

A simple model taking into account the backreaction due to the flux of negative energy particles through the horizons has been analyzed. It has been found that the black hole evaporates until asymptotically approaching a critical mass. In this way, the surface gravity of both horizons, their assigned temperature and the flow of energy of the tunneled particles decay towards a null value in the process of evaporation.

A preliminary test of stability has been performed on the inner horizon and it has been argued that stability under incoming radiation is only possible if the density of radiation decays as $u^{-p}$ with $p\geq 4$ and the surface gravity approaches its null value fast enough. Since the final black hole remnant has null surface gravity ($\kappa(u=\infty)=0$) and the classical radiative tail found by Price \cite{Price} follows a power-law with $L_{Ext.}(u)\sim u^{-p}$, where $p\geq 12$, it could be argued that the inner horizon is stable under the influence of the exterior incoming radiation\footnote{In fact, this is the argument that led to the conclusion of stability under incoming radiation of the extremal Reissner-Nordstr\"{o}m ($M=|Q|,\ \kappa=0$) black hole \cite{Poisson}.}.

In our treatment, it has been shown that the radiation coming from the tunneling in the inner horizon satisfies a power law in which $L_{IH}\sim u^{-3}$ so that a self-induced or \emph{endogenous instability} seems unavoidable independently of the decay followed by the surface gravity. Moreover, we have seen that just the ingoing flow of negative energy radiation coming from the outer horizon would have been enough to produce the endogenous instability on the inner horizon.


\end{document}